\newcommand\apj{{ApJ\,}}%
\newcommand\aap{{A\&A\,}}%
\newcommand\actaa{{Acta Astron.\,}}%
\documentclass[graybox, natbib, footinfo]{svmult}

\usepackage{mathptmx}       % selects Times Roman as basic font
\usepackage{helvet}         % selects Helvetica as sans-serif font
\usepackage{courier}        % selects Courier as typewriter font
\usepackage{type1cm}        % activate if the above 3 fonts are
\usepackage{makeidx}         % allows index generation
\usepackage{graphicx}        % standard LaTeX graphics tool
\usepackage{multicol}        % used for the two-column index
\usepackage[bottom]{footmisc}% places footnotes at page bottom
\usepackage{txfonts}
\usepackage{textcomp}

\makeindex             % used for the subject index
\begin{document}

\title*{The Metallicity Gradient of the Old Galactic Bulge Population}
\author{Sara Alejandra Sans Fuentes and Joris De Ridder}
\institute{S. Alejandra Sans Fuentes \at KU Leuven, Instituut voor 
Sterrenkunde, Celestijnenlaan 200D, B-3001 Leuven, Belgium 
\email{Alejandra.Sans@ster.kuleuven.be}
\and Joris De Ridder \at  KU Leuven, Instituut voor 
Sterrenkunde, Celestijnenlaan  200D, B-3001 Leuven, Belgium}
\maketitle

\abstract{Understanding the structure, formation and evolution of the Galactic 
Bulge requires the proper determination of spatial metallicity gradients in 
both the radial and vertical directions. RR Lyrae pulsators, known to be 
excellent distance indicators, may hold the key to determining these gradients. 
\cite{Jurcsik1996} has shown that RR Lyrae light curves and the phase 
difference of their Fourier decomposition, $\phi_{31}$, can be used to 
estimate photometric metallicities. The existence of galactic bulge 
metallicity gradients is a currently debated topic that would help pinpoint the 
Galaxy's formation and evolution. A recent study of the OGLE-III Galactic Bulge 
RR Lyrae Population by \cite{Pietrukowicz2012} suggests that the spatial 
distribution is uniform. We investigate how small a gradient would be 
detectable within the current S/N levels of the present data set, given the 
random and systematic errors associated with the derivation of a photometric 
metallicity versus spatial position relationship.}

\section{Introduction}
\label{sec:1}
The eleventh part of the Optical Gravitational Lensing Experiment (OGLE-III) 
contains the most complete and up-to-date photometric data set of Galactic 
Bulge RR Lyrae pulsating stars currently available. The sample contains 
photometric light curves in the I and V bands for a total of 11,756 fundamental 
mode RR Lyrae (RRab) \citep{Soszynski2011,Udalski2008}. This extensive data set 
has proven useful in investigating the Galactic interstellar extinction, 
metallicity gradients, and 3-D spatial 
distributions \citep{Pietrukowicz2012,Nataf2013,Smolec2005}. The work presented 
here uses 10,456 RRab stars from the OGLE-III Galactic Bulge RR Lyrae
catalog to investigate the detection limits of galactic bulge metallicity 
gradients.

\section{Methodology}
\label{sec:2}

The present sample includes RRab stars with I magnitudes between 
$13+\left(V-I\right)$ and $16+\left(V-I\right)$, in order to 
exclude foreground and background stars. Background stars are likely to 
belong to the Sgr dSph Galaxy \cite{Soszynski2011,Pietrukowicz2012}. From this 
sample, we create 1,000 population samples by assuming Gaussian distributions 
centered on the observed OGLE values and standard deviations equal to the 
respective errors for each star's Period, $\phi_{31}$, I and V. For each of 
these population samples, the photometric metallicities are estimated using the 
P-$\phi_{31}$-[Fe/H] relationship given in \cite{Smolec2005}: 
\begin{equation}
{\rm [Fe/H]}=-3.142-4.903P+0.824(\phi_{31} + \pi),
\end{equation}
where a phase shift is necessary due to sine/cosine fitting differences in 
\cite{Smolec2005} and \cite{Soszynski2011}. The galactocentric distances for 
each sample are calculated following the methodology presented in 
\citep{Pietrukowicz2012}:
\begin{eqnarray}
  \ \mathrm{log}\ Z={\rm [Fe/H]}-1.765\\
 M_{V}=2.288+0.882\ \mathrm{log}\ Z  +0.108(\mathrm{log}\ Z)^{2}\\
 M_{I}=0.471-1.132\ \mathrm{log}\ P  +0.025\ \mathrm{log}\ Z\\
 R_{I}=A_{I}/E(V-I)\\
 \ \mathrm{log}\ R= 1+0.2\left(I_{0}-M_{I}\right)\\
 R_{GCD} =\left[R_{0}^{2}+R^{2}-2RR_{0}\mathrm{cos}(l)\right]^{0.5}.
\end{eqnarray}
We estimate the mean sample barycenter to be 
$R_{0}=8.48 \pm 0.01$ Kpc and a mean sample metallicity of -0.994 $\pm$ 0.327 
for the OGLE-III Galactic Bulge RR Lyrae magnitude limited sample.

The effects that introducing a small synthetic gradient in the metallicity 
versus galactocentric distance plane would have on our observed variables are 
tested by  introducing a $\Delta$ [Fe/H] to each stars metallicity
based on its position.  A backwards propagation of the 
$\Delta$ [Fe/H] perturbation is performed, under the assumptions that
\begin{itemize}
 \item The distance to each star remains fixed,
 \item The ratio of total to selective extinction, $R_{I}$, remains unchanged,
 \item In cases of degeneracies, the induced perturbations are equally 
distributed between the fractional perturbations of all involved variables, 
\end{itemize}

A synthetic data set is created for which new values of 
[Fe/H] and $R_{GCD}$ and associated error distributions are 
calculated. In the case where $\Delta$ [Fe/H] values are introduced 
to create a synthetic gradient of -0.01 [dex/Kpc], our mean sample 
barycenter and mean sample metallicity become $R_{0}=8.47 \pm 0.01$ Kpc and 
[Fe/H]=-0.992 $\pm$ 0.329. Sampling through the defined [Fe/H] and $R_{GCD}$ 
distributions of each star, 1,000 least-squares (LS) fits are performed.  An 
F-test, to a confidence level of 0.05, is performed on each LS fit to determine 
 the significance of derived trends. The mean 
fit coefficients and their standard deviations are obtained for both 
samples.

\section{Preliminary Results}
\label{sec:3}
Re-deriving the OGLE-III Galactic Bulge RR Lyrae metallicities with the
latest methodology, and fitting the $R_{GCD}$ - [Fe/H] plane suggests that there 
exists a small but statistically significant galactic bulge metallicity 
gradient within the OGLE-III Galactic Bulge RR Lyrae database of strength 
-0.0239 $\pm$ 0.0006 [dex/Kpc]. In addition, using simulated data with an 
artificial trend of -0.01 [dex/Kpc] and realistic noise, a least squares fit 
results in an also statistically significant metallicity gradient of -0.012 
$\pm$ 0.0006 [dex/Kpc]. In both cases the significance of the gradient is tested 
to a confidence level of 0.05. Therefore, we conclude that with the current 
methodology and data sample, a gradient as small as -0.01 [dex/Kpc] can be 
detected within the OGLE-III Galactic Bulge RR Lyrae.

\begin{acknowledgement}
The authors would like to thank Igor Soszynski and Marcio Catelan for 
providing necessary fit and calibrations errors. This work was carried out, in 
whole or in part through the Gaia Research for European Astronomy Training 
(GREAT-ITN) network. The research leading to these results has received funding 
from the European Union Seventh Framework Programme ([FP7/2007-2013]) under 
grant agreement n\textdegree -264895.
\end{acknowledgement}
%

%\bibliographystyle{../aa.bst}
%\bibliography{references}

\begin{thebibliography}{6}
\expandafter\ifx\csname natexlab\endcsname\relax\def\natexlab#1{#1}\fi

\bibitem[{{Jurcsik} \& {Kovacs}(1996)}]{Jurcsik1996}
{Jurcsik}, J. \& {Kovacs}, G. 1996, \aap, 312, 111

\bibitem[{{Nataf} {et~al.}(2013){Nataf}, {Gould}, {Fouqu{\'e}}, {Gonzalez},
  {Johnson}, {Skowron}, {Udalski}, {Szyma{\'n}ski}, {Kubiak},
  {Pietrzy{\'n}ski}, {Soszy{\'n}ski}, {Ulaczyk}, {Wyrzykowski}, \&
  {Poleski}}]{Nataf2013}
{Nataf}, D.~M., {Gould}, A., {Fouqu{\'e}}, P., {et~al.} 2013, \apj, 769, 88

\bibitem[{{Pietrukowicz} {et~al.}(2012){Pietrukowicz}, {Udalski},
  {Soszy{\'n}ski}, {Nataf}, {Wyrzykowski}, {Poleski}, {Koz{\l}owski},
  {Szyma{\'n}ski}, {Kubiak}, {Pietrzy{\'n}ski}, \&
  {Ulaczyk}}]{Pietrukowicz2012}
{Pietrukowicz}, P., {Udalski}, A., {Soszy{\'n}ski}, I., {et~al.} 2012, \apj,
  750, 169

\bibitem[{{Smolec}(2005)}]{Smolec2005}
{Smolec}, R. 2005, \actaa, 55, 59

\bibitem[{{Soszy{\'n}ski} {et~al.}(2011){Soszy{\'n}ski}, {Dziembowski},
  {Udalski}, {Poleski}, {Szyma{\'n}ski}, {Kubiak}, {Pietrzy{\'n}ski},
  {Wyrzykowski}, {Ulaczyk}, {Koz{\l}owski}, \& {Pietrukowicz}}]{Soszynski2011}
{Soszy{\'n}ski}, I., {Dziembowski}, W.~A., {Udalski}, A., {et~al.} 2011,
  \actaa, 61, 1

\bibitem[{{Udalski} {et~al.}(2008){Udalski}, {Szymanski}, {Soszynski}, \&
  {Poleski}}]{Udalski2008}
{Udalski}, A., {Szymanski}, M.~K., {Soszynski}, I., \& {Poleski}, R. 2008,
  \actaa, 58, 69

\end{thebibliography}

\end{document}